\documentclass[aps,prl,twocolumn,groupedaddress,showpacs,amsmath,amssymb]{revtex4}
\usepackage{epsfig}
\usepackage{amssymb}
\usepackage{amsmath}
\usepackage{graphicx}
\usepackage{amsfonts}
\usepackage{epsfig}
\usepackage{revsymb}
\usepackage{bm}
\usepackage{latexsym}
\usepackage{hyperref}

\begin{document}

\title{Rayleigh-Taylor instability of crystallization waves at the superfluid-solid $^4$He interface}

\author{S. N. Burmistrov}
\email[]{burmi@kurm.polyn.kiae.su} \affiliation{Institute of Superconductivity and Solid State Physics,
Kurchatov Institute, 123182 Moscow, Russia}
\author{L. B. Dubovskii}
\affiliation{Institute of Superconductivity and Solid State Physics, Kurchatov Institute, 123182 Moscow,
Russia}
\author{V. L. Tsymbalenko}
\affiliation{Institute of Superconductivity and Solid State Physics, Kurchatov Institute, 123182 Moscow,
Russia}

%\date{\today}

\begin{abstract}
At the superfluid-solid $^4$He interface there exist crystallization waves having much in common with
gravitational-capillary waves at the interface between two normal fluids. The Rayleigh-Taylor instability is
an instability of the interface which can be realized when the lighter fluid is propelling the heavier one. We
investigate here the analogues of the Rayleigh-Taylor instability for the superfluid-solid $^4$He interface.
In the case of a uniformly accelerated interface the instability occurs only for a growing solid phase when
the magnitude of the acceleration exceeds some critical value independent of the surface stiffness. For the
Richtmyer-Meshkov limiting case of an impulsively accelerated interface, the onset of instability does not
depend on the sign of the interface acceleration. In both cases the effect of crystallization wave damping is
to reduce the perturbation growth-rate of the Taylor unstable interface.
\end{abstract}


\maketitle

\section{INTRODUCTION}
\par
The Rayleigh-Taylor instability is a fingering instability of an
interface between two fluids of different densities. It takes place when the
heavier fluid is decelerated by the lighter fluid or, in other words, density
and pressure gradients have opposite directions \cite{T,Ch}. The
Rayleigh-Taylor instability \cite{Sh} occurs in numerous physical and
technological situations, e.g., gravity-driven instability of a heavier fluid
atop a lighter one and inertial confinement fusion. In essence, the
Rayleigh-Taylor instability is the first step in a fluid-mixing mechanism, the
step eventually leading via formation of bubbles, spikes, and curtains to the
turbulent regime of fluid mixing. Along with the Kelvin-Helmholtz criterion
for tangential discontinuities at the interface between two normal fluids,
the Rayleigh-Taylor criterion is among the most generic principles in the
complicated subject of interface instability.
\par
For the interfaces of superfluid $^4$He, some of the hydrodynamic instabilities have been found as well.
First, we mention the Faraday instability which denotes the phenomenon of the parametric excitation of
standing waves on the free surface of a fluid. The flat shape of the surface becomes unstable with a periodic
modulation of the acceleration of gravity. Recently \cite{Ab}, generation of Faraday standing waves on the
free surface of $^4$He has been realized in the experimental cell subjected to sinusoidal vibration in the
vertical direction. According to \cite{Ki,Kim}, it is also possible to generate a dense fog of helium droplets
by driving the capillary waves on a superfluid $^4$He surface unstable with an intense ultrasonic beam from a
piezoelectric transducer under the surface. There have been observed electrohydrodynamic interface
instabilities due to charges trapped at the surfaces and interfaces of various condensed helium phases
\cite{Le}. The shear flow between the superfluid \textit{A} and \textit{B} phases of $^3$He can result in the
Kelvin-Helmholtz interface instability \cite{Bl}.
\par
Dynamics of the superfluid-solid $^4$He interface due to sufficiently fast processes of crystallization and
melting resembles much that of the free surface of a fluid. In particular, as was predicted by Andreev and
Parshin, the crystal in contact with its liquid phase can support wave-like processes of crystallization and
melting, see review \cite{BAP}. From the dynamical point of view such weakly damping crystallization waves are
an immediate counterpart of the well-known gravitational-capillary waves at the vapor-liquid interfaces.
\par
A series of mechanical and hydrodynamical instabilities has been predicted and
observed for the superfluid-solid $^4$He interface. We mention the Grinfeld
instability under uniaxial stress of a solid. Warping of the flat interface
occurs at some threshold stress when the release of elastic energy exceeds the
loss of the surface energy \cite{No,BAP}. Like normal fluids, the steady flow
of a superfluid in the direction tangential to the interface can result in the
Kelvin-Helmholtz instability. As the flow exceeds a threshold magnitude,
crystallization waves appear at the superfluid-solid $^4$He interface
\cite{Ka,Noz}. The phenomenon has qualitatively been observed as a distortion
of the crystal surface in the fluid jet \cite{Ma}. An analogy with generating
sea waves by wind is fully appropriate here.
\par
To date, the Rayleigh-Taylor phenomena have extensively been studied in normal fluids, but not much study has
been made in superfluids or quantum solids. The classical Rayleigh-Taylor instability of the superfluid-solid
$^4$He interface in the field of gravity is observed by Demaria, Lewellen, and Dahm \cite{De}. In these
experiments a cell in which the solid and liquid phases occupy initially the lower and upper halves,
respectively, is inverted mechanically by 180$^{\circ}$. After inversion a single finger of the liquid phase
ascends at the centre of a cell, and the solid phase descends along the walls. On the other hand, in
experiment \cite{Ts} the flat shape of the interface remained stable for a $^4$He crystal grown at the needle
point with its lower facet under favorable condition for developing the gravity-driven instability. A
difference in the observations can be associated with the following reasons. The requirement for the interface
instability as well as the initial stage of fingering process, as is shown in \cite{Bu}, is sensitive to the
state of a crystal facet, rough or smooth, and to the size of a facet.
\par
Recently \cite{Abe}, it has been demonstrated that the Faraday instability is also inherent in the
superfluid-solid $^4$He interface. Crystallization waves at the horizontal interface between superfluid and
solid $^4$He are generated by a periodic vibration of an experimental cell in the vertical direction. In
accordance with expectation \cite{Sa} the amplitude of the waves excited at one-half of the driving frequency
decreases for higher temperatures due to reduction of the interface growth coefficient describing dissipative
properties of the interface. From general arguments the Faraday instability can be viewed as a particular case
of the Taylor instability for the periodically driven interface.
\par
\begin{figure}
\includegraphics[scale=2.2]{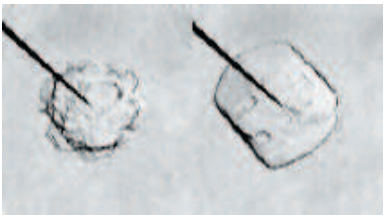}
\caption{The growth of a $^4$He crystal at 0.47~K and initial overpressure 5.2~mbar. The left frame
corresponds to 0.19~ms after nucleation. The right frame is taken 80~ms later when the net pressure is already
close to the melting pressure. The vertical size of the frames is 2.4~mm.}
\label{fig1}
\end{figure}

On the other side, the spectrum of crystallization waves remains invariable for the steady flow of a
superfluid in the direction normal to the interface \cite{Noz}. At first sight this implies that the growth of
a solid should not bring the superfluid-solid interface to instability. However, in the experiments on free
growth of a $^4$He crystal initiated at the needle point immersed into the overpressurized liquid bulk
\cite{Tsy,Tsym} one can observe a destruction of the regular shape of the crystal triggered under sufficiently
large overpressure exceeding about 6~mbar. Immediately after nucleation the crystal seed has a clear hexagonal
prism-like shape with slight ripples. Soon afterwards by 0.1--0.2~ms the shape of the crystal becomes round
with a highly irregular outline. Far later by 100--150~ms, as the net overpressure vanishes and the pressure
in the cell becomes phase-equilibrium, the shape of the grown crystal relaxes to a regular hexagonal prism. We
put two images of a crystal in Fig.~\ref{fig1}.
\par
\begin{figure}
\includegraphics[scale=0.8]{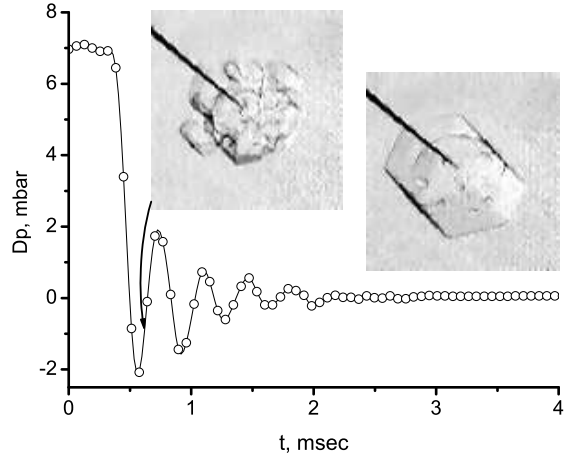}
\caption{The deviation of the pressure from equilibrium vs time during crystal growth at 0.48~K and initial
overpressure 7~mbar. The left insert shows a crystal at 0.64~ms after nucleation. The time at the right insert
is 16~ms. The vertical size of the frames is 2.4~mm.} \label{fig2}
\end{figure}

For overpressures higher than 6~mbar, see Fig.~\ref{fig2}, we discover more exotic patterns by the same time
interval 0.1--0.2~ms after the voltage pulse which triggers nucleation. The interfacial irregularities become
more pronounced and acquire a mushroom-like shape. Apparently, the fluid moves into the crystal between the
neighboring spikes, resulting eventually in generation of liquid bubbles inside the crystal.
\par
In addition, in Fig.~\ref{fig2} we show the behavior of the pressure in the course of the crystal growth.
After nucleation of a crystal seed the overpressure in the cell drops and then gradually vanishes, oscillating
around zero value corresponding to the equilibrium pressure. The pressure oscillations are due, in the first
turn, to the finiteness of the experimental volume and the finiteness of sound velocity. The point is that the
appearance of a solid seed and its next growth are accompanied by variations of the density and the volume in
which the density changes. The change of the seed volume results in exciting and emitting the sound waves
which propagate in the direction to the container walls with the next backward reflection from the walls to
the solid seed.  E.g., the first- and second-sound emissions with an expanding $^3$He-concentrated drop in a
superfluid $^3$He-$^4$He mixture has been analyzed in Ref.~\cite{BS}. The excitation and emission of sound
become more and more effective as the interface rate and acceleration increase. Eventually, we obtain acoustic
damping oscillations of the liquid inside the cell~\cite{Tsymb}. In some sense the damping of the pressure
oscillations represents a quality factor of a liquid/solid or melting/freezing resonator. In the lack of any
dissipative processes in the system the pressure oscillations will last infinitely long.
\par
The pressure in the cell becomes equilibrium pressure in 2~ms and the driving force vanishes. Finally at
16~ms, the crystal relaxed and acquired the regular hexagonal shape. However, we still observe the liquid
bubbles embedded into the crystal bulk.
\par
If we roughly estimate the velocity which the interface should acquire by the time of the overpressure release
and formation of the irregular outline, we will find a rather high magnitude of several meters per second. The
corresponding acceleration which provides such increment of velocity should be about 10$^3g$, $g$ being
acceleration of gravity. On the whole, large acceleration for a short time shows evidence in favor of a
shock-driven character of the crystal growth. It is interesting to note that the irregular patterns observed
are similar in appearance to those obtained in studies \cite{Me} of a shock-accelerated boundary between two
gases of different densities. The typical time of the pattern formation was of the same order of several
tenths of millisecond.
\par
In the present work we develop a linear theory on the Taylor instability of an arbitrarily accelerated
boundary between the superfluid and solid $^4$He phases. In essence, we derive an equation which the small
interfacial perturbations obey. We consider three typical cases of the interface acceleration: constant,
shock, and periodic. The plane and  spherical interface geometries are analyzed.

\section{PLANE GEOMETRY AND THE INTERFACE GROWTH KINETICS}
\par
Let us assume the interface is parallel to the $x$-$y$ plane, with vertical position $z=L(t)$ which moves at
the rate $V=\dot{L}(t)$. The upper half-space $z>L(t)$ is occupied with the liquid phase and the solid phase
occupies the lower one $z<L(t)$. Below we will consider the stability of the moving interface with respect to
its small perturbations $\zeta (x, y,\, t)$ from the flat shape. Thus, the coordinate $Z(x, y,\, t)=L(t)
+\zeta (x, y,\, t)$ gives the vertical position of the perturbed interface evolving in time.
\par
To discover the effect of nonuniform motion on the stability of the interfacial shape, we make a number of
simplifying assumptions which do not affect the main point of the phenomenon. The validity and criteria of
applicability for the assumptions to be made below can be found in the papers on the kinetic interface
coefficients and crystallization waves~\cite{CN,BE,AK,UB}.
\par
So, in the superfluid we employ the usual two-fluid equations without dissipation. In addition, we assume that
the growth rate of a solid $V(t)$ is always small compared with the velocity of the first or second sound.
This is an ordinary experimental situation. Thus we treat the hydrodynamics of the superfluid in the
approximation of incompressible liquid and the constancy of the entropy density. In this case \cite{LL} the
velocities $\bm{v}_n$ and $\bm{v}_s$ of the normal and superfluid motions can be described in terms of
gradient of velocity potentials $\phi _n$ and $\phi _s$ which satisfy $\nabla^2 \phi _n=0$ and $\nabla^2 \phi
_s=0$, respectively. Since finally we will discuss only the linearized equations in  the perturbation $\zeta$,
it is convenient to consider a single Fourier mode of the perturbation $\zeta =\zeta _q (t)\exp
(i\bm{q}\bm{r})$ with wave vector $\bm{q}=(q_x,\, q_y)$ parallel to the boundary. The solutions of $\nabla^2
\phi _{n,\, s}=0$ can be represented as
\begin{gather}\nonumber
\phi _s=u_s(t)z+A_s(t)\exp (i\bm{q}\bm{r}-qz) ,
\\
\phi _n=u_n(t)z+A_n(t)\exp (i\bm{q}\bm{r}-qz) , \label{f01}
\end{gather}
where $\bm{r}=(x,y)$ and the velocities $u_s(t)$ and $u_n(t)$ describe the
undisturbed motion in the superfluid. The perturbation amplitudes $A_s(t)$ and
$A_n(t)$ are assumed to be linear in $\zeta _q(t)$ and will be determined
later from the corresponding boundary conditions at the interface. The
pressure in the superfluid is a sum of pressures $P=P_n+P_s$ and
\begin{gather}\nonumber
P_s=P_{s\,\infty}-\rho _s\bigl(\dot{\phi}_s+(\nabla\phi _s)^2/2\bigr)-\rho _sgz ,
\\
P_n=P_{n\,\infty}-\rho _n\bigl(\dot{\phi}_n+(\nabla\phi _n)^2/2\bigr)-\rho _ngz . \label{f02}
\end{gather}
The index \textquotedblleft $\infty$\textquotedblright\ refers to the values taken away at infinity.
\par
Unlike previous considerations \cite{CN,BE,AK,UB} which are also linear in the interfacial perturbations, we
have to retain the quadratic terms in the superfluid and normal velocities $\bm{v}_{s,\, n}=\nabla\phi _{s,\,
n}$ on account of nonzero value $u_{s,\, n}(t)$ and product like $u_sA_s$ or $u_nA_n$.
\par
The mass continuity across the boundary gives at $z=Z(x, y,\, t)$
\begin{equation}\label{f03} j_{\nu}=\rho _nv_{n\,\nu} +\rho _sv_{s\,\nu}
=(\rho -\rho ')\dot{Z}\, ,
\end{equation}
where $\dot{Z}$ is the velocity of
the boundary, $\bm{\nu}$ is the unit vector normal to the boundary, and
$j_{\nu}$ is the mass current normal to the boundary. The densities $\rho _n$,
$\rho _s$ are the normal and superfluid densities, $\rho =\rho _n+\rho _s$,
and $\rho '$ is the density of the solid phase. The normal components of
velocities can be approximated by $v_{n\,\nu}\approx v_{n\, z}$ and
$v_{s\,\nu}\approx v_{s\, z}$. We also believe that there is no motion in the
solid phase, i.e., $v'=0$.
\par
To further simplifications, we suppose the normal component sticks to the interface like a viscid fluid. Also,
this implies the Kapitza resistance to be infinite \cite{BE}. In addition, we disregard any excitations, e.g.
phonons, in the solid. So, we put at the boundary
\begin{equation}\label{f04}
v_{n\,\nu}=\dot{Z}\, .
\end{equation}
Using Eqs.~(\ref{f01}), (\ref{f03}), and (\ref{f04}), we can determine the
unknown amplitudes in (\ref{f01}) and then the velocity fields $\bm{v}_s$ and
$\bm{v}_n$ in the liquid. For the unperturbed motion, it is obvious
\begin{equation}\label{f05}
u_s=-\,\frac{\rho '-\rho _s}{\rho _s}\, V \;\;\text{and}\;\; u_n=V\, .
\end{equation}
The amplitudes $A_n(t)$ and $A_s(t)$ are given approximately by
\begin{gather}\nonumber
A_s(t)=\frac{\rho '-\rho _s}{\rho _s}\,\frac{\dot{\zeta}_q(t)}{q}\, \text{e}^{\, qL(t)}\, ,
\\
A_n(t)=-\,\frac{\dot{\zeta}_q(t)}{q}\, \text{e}^{\, qL(t)}\, .\nonumber
\end{gather}
As a result, we can also calculate the pressure field (\ref{f02}) in the
liquid. The boundary conditions (\ref{f03}) and (\ref{f04}) in combination
with the obvious relations (\ref{f05}) determine unambiguously the inertial
properties of the interface described in terms of the effective density
\cite{BE}
\begin{eqnarray*}
\rho _{\text{ef}}=\rho _n +(\rho '-\rho _s)^2/\rho _s
\end{eqnarray*}
in the sense that
\begin{eqnarray*}
\rho _su_s^2/2 + \rho _nu_n^2/2 = \rho _{\text{ef}}V^2/2\, .
\end{eqnarray*}
\par
To proceed further, we adopt the most simplifying assumptions \cite{BE} to describe the solid and its
boundary. The solid is assumed to be always unstressed and all possible shearing components $\sigma _{i\neq
k}$ of the stress tensor $\sigma _{ik}$ are neglected. In other words, the stress tensor is isotropic, i.e.,
$\sigma _{ik}=-P'\delta _{ik}$, and we can define \textquoteleft pressure\textquoteright\ according to
$P'=-\sigma _{ii}/3$ \cite{UB}. Then, from the formal point of view, the solid can be described as a liquid
under pressure equal to $P'$.
\par
The next boundary condition stems from the continuity of the momentum flux density across the interface. The
momentum flux density in the superfluid \cite{LL} reads $P\delta _{ik}+\rho _nv_{n\, i}v_{n\, k}+\rho _sv_{s\,
i}v_{s\, k}$. Then, we take $\sigma _{ik}\nu _k=-P'\nu _i$ into account, assume the small curvature of the
interface $z=Z(x, y,\, t)=L(t)+\zeta (x, y,\, t)$, and use a frame that refers to the boundary
\begin{gather}\nonumber
P+\rho _n(\bm{v}_n-\dot{Z}\bm{\nu})^2+\rho _s(\bm{v}_s-\dot{Z}\bm{\nu})^2-(P'+\rho '\dot{Z}^2)=
\\
 =\gamma _{ik}\partial Z ^2/\partial r_i\partial r_k
  =\gamma _{ik}\partial\zeta ^2/\partial r_i\partial r_k\, . \label{f06}
\end{gather}
Neglecting the quadratic terms in velocities gives the usual Laplace condition of mechanical equilibrium
across the interface \cite{BE}. Here $\gamma _{ik}(\theta ,\varphi )=\alpha\delta _{ik} +\partial\alpha
^2/\partial\varphi _i\partial\varphi _k$ is the surface stiffness tensor \cite{BAP,No,BE} expressed in terms
of surface tension $\alpha =\alpha (\theta ,\varphi )$ depending on the angles between the crystalline
orientation and the normal to the surface.
\par
Let us turn now to the last boundary condition. It is a reasonable assumption that any motion of the interface
accompanied also by the melting and growth of a solid will dissipate a certain amount of energy. Thus a finite
velocity of the interface should produce some imbalance in the chemical potential difference $\mu -\mu '$
between the liquid and solid. The routine in various theories of the interfacial dynamics is an introduction
of the so-called growth coefficient $K$ which relates the interface growth rate with the difference in
chemical potentials across the interface \cite{CN,BE}. Because of $u_n(t)\neq 0$ and $u_s(t)\neq 0$ we again
have to take into account the squares of velocities which are always omitted in the linear perturbation theory
of the interface being initially at rest. So, at the boundary we employ an effective relation
\begin{gather}\label{f07}
\dot{Z}=K\left[\mu +\frac{(\bm{v}_s-\dot{Z}\bm{\nu})^2}{2}-\left(\mu '+\frac{\dot{Z}^2}{2}\right)\right]\, ,
\end{gather}
where $\mu$ and $\mu '$ are the chemical potentials of the liquid and solid
per unit mass. In a wide sense the growth coefficient here is a certain
combination of all Onsager coefficients and the kinetic coefficients
describing the near-surface dissipative processes. In general, the growth
coefficient $K$ can depend on the temperature as well as on the wave vector
$q$. Usually, in the ballistic regime, when the mean free path $l$ of
excitations is large, the growth coefficient is independent of wave vector. In
the opposite hydrodynamic limit $ql\ll 1$ the growth coefficient may depend on
the wave vector approximately as $1/K\sim ql$ \cite{BE}.
\par
Lastly, we need an expression for the chemical potential difference. As usual, the reference point is the
melting pressure $P_c$ at which the chemical potentials $\mu$ and $\mu '$ coincide and the liquid-solid
transition takes place. We take the necessary formulae for the superfluid from Ref.~\cite{LL}. After expanding
chemical potentials in the vicinity of the melting pressure, we obtain
\begin{gather}
\;\;\;\;\;\;\; \mu -\mu ' = \nonumber
 \\ =  \sigma (T-T_{\infty})+\frac{P-P_c}{\rho}-\frac{\rho
_n}{\rho}\,\frac{(\bm{v}_n-\bm{v}_s)^2}{2}-\,\frac{P'-P_c}{\rho '}\, , \nonumber
\\
 T-T_{\infty}
 =\! \frac{\rho _n}{\sigma\rho}\left(\frac{P_n-P_{n\,\infty}}{\rho _n}-\frac{P_s-P_{s\,\infty}}{\rho
_s}-\frac{(\bm{v}_n-\bm{v}_s)^2}{2}\right) ,\nonumber
\end{gather}
where $\sigma$ is the entropy and the quantities with index \textquotedblleft
 $\infty$\textquotedblright\, stand for the magnitudes taken far from the interface.
\par
Now we are in position to find the equations which the interface dynamics
obeys. Knowing velocity potentials $\phi _n$ and $\phi _s$ expressed via
$u_{n,\, s}(t)$ and perturbation $\zeta _q(t)$, we can calculate the normal
and superfluid velocities, pressure, and chemical potential difference. Next,
we insert the quantities calculated at the interface into the boundary
conditions (\ref{f06}) and (\ref{f07}) and eliminate the pressure $P'$. As a
result of some algebraic formula manipulation linear in $\zeta _q$, we obtain
an equation consisting of the $\zeta _q$-independent component and the other
one linear in $\zeta _q$. The first component gives an equation
\begin{equation}
V\frac{\rho '}{K}=\frac{\rho '-\rho}{\rho}\, \bigl[\Delta P-\rho gL)\bigr] + \rho
_{\text{ef}}\!\left(\dot{V}L+\frac{V^2}{2}\right) ,\nonumber
\end{equation}
which describes the undisturbed motion of the flat interface and relates overpressure $\Delta
P(t)=P_{\infty}(t)-P_c$ to $V(t)=\dot{L}(t)$ in a complicated manner in order to support the necessary
behavior of the growth rate. This equation does not have much interest for us.
\par
The other equation obtained is the most significant one. It represents the equation for the linear dynamics of
the interface perturbation $\zeta =\zeta _q(t)\exp (i\bm{qr})$ when the interface is subjected to an arbitrary
driving acceleration $\dot{V}(t)$
\begin{equation}\label{f1}
\rho _{\text{ef}}\,\frac{\ddot{\zeta}_q}{q} +\frac{\rho '}{K}\,\dot{\zeta}_q  + \bigl[\gamma _{ik}q_iq_k
+(\rho '-\rho)g-\rho _{\text{ef}}\dot{V}(t)\bigr]\,\zeta _q =0 .
\end{equation}
The new aspect of the equation derived is an additional term with the interface acceleration $\dot{V}(t)$. As
is expected, the uniform motion of the interface at $V(t)=\text{const}$ does not influence the character of
small interfacial oscillations. For $\dot{V}(t)=0$, Eq.~(\ref{f1}) amounts to the known relation determining
the spectrum of crystallization waves when the melting-crystallization processes are balanced and the
interface position in average is invariable, i.e., $L(t)=\text{const}$ \cite{AK,BE,BAP}. \par Undoubtedly,
more realistic and complicated models of the superfluid-crystal $^4$He interface will improve the magnitudes
of the effective interface density, effective growth coefficient, and surface stiffness. However, we believe
that the structure of Eq.~(\ref{f1}) is generic and holds.

\section{ INTERFACE INSTABILITIES}
\par
Equation (\ref{f1}) can have unstable solutions depending strongly on the
acceleration history of the interface. First, let us consider the stability of
the plane interface with respect to small perturbations $\zeta _q\sim\exp
(\lambda (q)\, t)$ for the uniformly accelerated growth of a crystal. The root
with $\text{Re}\,\lambda (q)>0$ means the interface instability, i.e.,
initially small-amplitude perturbations of wavelength $2\pi /q$ will grow
exponentially in time. For the acceleration exceeding the threshold
$\dot{V}_c=g(\rho '-\rho)/\rho _{\text{ef}}$, the interfacial perturbation
will increase for the wave vectors satisfying $\gamma q_c^2<\rho
_{\text{ef}}\dot{V}-(\rho '-\rho)g$. For brevity, we put $\gamma _{ik} q_iq_k
= \gamma q^2$ where $\gamma =\gamma _{ik}n_{i}n_k$ and $\bm{n}=\bm{q}/q$ is
unit vector in the direction of perturbation. Note that the threshold
acceleration $\dot{V}_c$ does not depend on the growth coefficient, i.e., on
the dissipative properties of the interface, and is positive. The latter
corresponds to the case when the interfacial acceleration $\dot{V}$ is
directed to the superfluid. Thus, the Taylor instability due to non-uniform
growth of the plane interface appears only for the accelerated growth of a
solid.
\par
The surface stiffness stabilizes the region of long wave perturbations and establishes the most unstable
wavelength having the fastest exponential growth. The value $q_0$ corresponding to the maximum magnitude
$\text{Re}\,\lambda (q)$ gives the shortest time scale for the development of the instability which will be
characterized by the wavelength $2\pi /q_0$. The value $q_0$ can be found from the equation
\begin{eqnarray*}
q_0^2=\frac{q_c^2}{3}-\,\frac{\sqrt{2}}{3}\,\frac{\rho '}{\sqrt{\gamma\rho
_{\text{ef}}}}\,\frac{q_0^{3/2}}{K}\, ,
\end{eqnarray*}
where $q_c$ is the value related to the upper bounds of instability according to $\gamma q_c^2=\rho
_{\text{ef}}\dot{V}-g(\rho '-\rho)$. For large magnitudes of the growth coefficient or large acceleration in
the case $K^4\dot{V}\gg\rho ^{\prime\, 4}/\gamma\rho _{\text{ef}}^3\,$, the values $q_0$ and $\lambda (q_0)$
are approximately equal  to
\begin{gather*}
q_0=\frac{q_c}{\sqrt{3}}\left(1-\,\frac{\rho '}{(6\sqrt{3}\,\gamma\rho _{\text{ef}}\,K^2q_c)^{1/2}}\right) ,
\\
\lambda _0=\left(\frac{2\gamma}{3\sqrt{3}}\,\frac{q_c^3}{\rho _{\text{ef}}}\right)^{1/2}
-\,\frac{q_c}{2\sqrt{3}}\,\frac{\rho '}{\rho _{\text{ef}}\, K}\,  .
\end{gather*}
The values $q_0$ and $\lambda _0$ depend on $\gamma$, thus implying anisotropic and complicated possible
surface patterns.
\par
In the opposite limit $K^4\dot{V}\ll\rho ^{\prime\, 4}/\gamma\rho _{\text{ef}}^3$ one has roughly
\begin{gather*}
q_0=\left(\frac{\gamma\rho _{\text{ef}}}{\rho ^{\prime\, 2}}\, K^2q_c^4\right)^{\! 1/3}  ,
\\
\lambda _0=\frac{K}{\rho '}\,\gamma q_c^2 = \frac{K}{\rho '}\bigl ( \rho _{\text{ef}}\dot{V}-(\rho '-\rho
)g\bigr )\, .
\end{gather*}
It is interesting that, though the spatial scale $q_0^{-1}$ is sensitive to the surface stiffness and its
anisotropy, the time of developing the instability becomes independent of the surface stiffness and its
anisotropy.
\par
On the whole, the values $q_0$ and $\lambda _0$ decrease as the kinetic growth coefficient reduces or
dissipation with the interface enhances. From the experimental point of view this may require a crystal
surface of sufficiently large sizes $d>2\pi /q_0$ and a large time to support the accelerated growth
$t>1/\lambda _0$ in order to realize an interfacial instability with a uniformly accelerated growth.
\par
In Ref.~\cite{Sa} it has been shown that periodic modulation of the gravitational constant as
$g(t)=g(1+2\tilde{\epsilon}\cos 2\omega t)$ can result in parametric excitation of crystallization waves at
the stationary flat interface corresponding to $V(t)=0$ in our case. Admitting some analogy between gravity
and noninertial frame, one can expect a possibility of exciting crystallization waves with a periodic driving,
e.g., $\dot{V}(t)=G\cos 2\omega t$, even with the lack of gravity. To demonstrate this, let us rewrite
Eq.~(\ref{f1}) in the form
\begin{equation}\label{f2}
\ddot{\zeta}_q+\Gamma _q\dot{\zeta}_q+\omega _0^2(q)\bigl(1-\dot{V}(t)/\tilde{g}\bigr)\,\zeta _q =0 ,
\end{equation}
where we have introduced the damping coefficient $\Gamma _q$, the frequency $\omega _0(q)$ of crystallization
waves in the lack of damping, and the scaled accelerating amplitude $\tilde{g}$
\begin{gather*}
\omega ^2_0(q)=\bigl(\gamma _{ik}q_iq_k+(\rho '-\rho)g\bigr)q/\rho _{\text{ef}}, \,\,\, \Gamma _q=\frac{\rho
'}{\rho _{\text{ef}}}\,\frac{q}{K}\; ,
\\
\tilde{g}=g\,\frac{(\rho '-\rho)}{\rho _{\text{ef}}}\left(1+\frac{\gamma _{ik}q_iq_k}{(\rho '-\rho)g}\right) .
\end{gather*}
Then, if a periodic process of melting and crystallization is realized in experiment so that the interface
could oscillate around some average position at frequency $2\omega$ and amplitude $G/(2\omega)^2$, equation
(\ref{f2}) with label $2\epsilon =-G/\tilde{g}$ transforms to a Mathieu-type equation
\begin{equation}\nonumber
\ddot{\zeta}_q+\Gamma _q\dot{\zeta}_q+\omega _0^2(q)\bigl(1+2\epsilon\cos 2\omega t\bigr)\,\zeta _q =0 ,
\end{equation}
which is identical to that analyzed in \cite{Sa}. The predictions which follow
are well known and we refer the readers to papers \cite{Sa,Abe} for details.
Note only that in the free crystal growth experiments \cite{Ts,Tsym} the
pressure in the cell drops drastically down after nucleating a solid seed with
the subsequent transition to the damped oscillations around the melting
pressure.
\par
Crystallization waves at the superfluid-solid $^4$He interface can also be generated with the
Richtmyer-Meshkov mechanism when the interface is subjected to an impulsive acceleration, i.e.,
$\dot{V}(t)\sim V(0)\delta (t)$. In ordinary fluids and gases, for this purpose, the passage of a shock wave
across the interface is commonly used. Unlike the Taylor case of constant acceleration when the perturbation
amplitude in the linear regime grows exponentially in time, the initial stage of interface instability in the
Richtmyer-Meshkov case of shock acceleration \cite{Ya} is characterized by a linear growth of the perturbation
amplitude in time. The Richtmyer-Meshkov instability is independent of the direction of acceleration in
contrast to the Taylor one. The late time stages of both instabilities may demonstrate a formal resemblance,
showing bubble and spike morphology.
\par
According to Eq.~(\ref{f1}), the growth rate of the Richtmyer-Meskov unstable
interface in the linear regime can approximately be described in the overdamped regime as
\begin{gather}
\zeta _q(x,\, t)=qV(0)\,\text{e}^{-\Gamma _qt/2}\nonumber
\\ 
\times 
\frac{\sinh\biggl(t\sqrt{\Gamma _q^2/4-\omega _0^2(q)}\biggr)}{\sqrt{\Gamma _q^2/4-\omega _0^2(q)}}\,\zeta _q(x\, ,0) ,\;\;\; \omega _0(q)< \Gamma _q/2 \, ;
\label{f3}
\end{gather}
and in the weakly damped regime as 
\begin{gather*}
\zeta _q(x,\, t)=qV(0)\,\text{e}^{-\Gamma _qt/2}\nonumber
\\
\times
\frac{\sin\biggl(t\sqrt{\omega _0^2(q)-\Gamma _q^2/4}\biggr)}{\sqrt{\omega _0^2(q)-\Gamma _q^2/4}}\,\zeta _q(x\, ,0) ,\;\;\; \omega _0(q)>\Gamma _q/2\, .
\end{gather*}
Here $\zeta _q(x\, ,0)=\zeta _q(0)\cos qx$ is the initial perturbation amplitude at the interface immediately
after the impulsive acceleration and $V(0)$ is an increment in the interface velocity due to acceleration of
the boundary. At the linear stage the amplitude of crests and troughs is approximately the same. The shape of
the crests and troughs is similar and the interface remains approximately sinusoidal. Staying in the linear
regime, we see from (\ref{f3}) that the finite damping of crystallization waves is a stabilizing factor and
can limit the growth of the perturbation amplitude. 
\par 
Next, we consider two extreme cases of $\omega _0(q)=0$ and $\Gamma _ q=0$. For $\omega _0(q)=0$ which corresponds mainly to the flat interface perturbations with small wave vectors $q$, we have
\begin{gather*}
\zeta _q(x,\, t)=\frac{qV(0)}{\Gamma _q}\,\biggl(1-\,\text{e}^{-\Gamma _qt}\biggr)\zeta _q(x\, ,0)
\\
=\frac{\rho _{\text{ef}}}{\rho '}\, K\, V(0)\,\biggl(1-\,\text{e}^{-\Gamma _qt}\biggr)\zeta _q(x\, ,0) , \;\;\;\omega _0(q)=0 \, .
\end{gather*}
Here it is worthwhile to note that the total amplification
factor is proportional to the interface growth coefficient $K$ and proves to be independent of the wavelength
$2\pi/q$ of an initially sinusoidal perturbation. On the contrary, the time for ceasing the growth of
perturbation amplitude is scale-dependent. The larger the wavelength, the longer the decay time. A special
feature of crystallization waves in $^4$He is that the growth coefficient $K$, and thus $\Gamma _q$, is
strongly temperature-dependent. 
\par 
For the opposite case when the crystallization wave damping is absent $\Gamma _q=0$, we arrive at the persistent oscillations of the initial interface perturbation
\begin{gather*}
\zeta _q(x,\, t)=qV(0)\,
\frac{\sin\bigl(t\omega _0(q)\bigr)}{\omega _0(q)}\,\zeta _q(x\, ,0) ,\;\;\; \Gamma _q=0\, .
\end{gather*}
The maximum amplitude of oscillations is governed by the crystallization frequency $\omega _0(q)$. Since roughly $\omega _0(q)\sim q^{3/2}$, the large curvature of the initial interface perturbation can also be a factor that limits the infinite growth of perturbation amplitudes.
\par 
In $^4$He the Richtmyer-Meshkov instability can be studied with a sound wave
pulse hitting the interface in the normal direction. Recently \cite{Nomu}, it has experimentally been
demonstrated that the superfluid-solid $^4$He interface can be set in motion with a sound wave which transmits
through the interface, giving rise to the processes of crystallization and melting.

\section{SPHERICAL GEOMETRY}
\par
In experiment \cite{Tsym} a solid nucleates at the needle point and then grows free. A ratio of the crystal
sizes in the different directions for the crystals grown is not drastic. The ratio of the maximum to minimum
size does not exceed 2--3, see also Figs.~\ref{fig1} and \ref{fig2}. Here we consider the stability of the
spherical shape of a growing solid. For simplicity, we assume the surface tension $\alpha$ is isotropic. In
addition, we neglect the acceleration of gravity and, as above, treat the equations linearized in the
interface perturbation.
\par
The equation specifying the interface motion is taken as $r=R_s(t,\, \Omega )=R(t)+\zeta (t,\, \Omega )$,
where $R(t)$ is the radius of the unperturbed spherical interface and $\zeta (t,\, \Omega )=\sum _l\zeta
_l(t)Y_l(\Omega )$ is an interface perturbation expanded in the spherical harmonics of degree $l=0,\,1,\,
2\ldots$ We look for velocity potentials of the normal $\bm{v}_n=\nabla\phi _n$ and superfluid
$\bm{v}_s=\nabla\phi _s$ motions of the form
\begin{gather*}
\phi _s =-u_s(t)R^2/r\, +A_l(t)Y_l/r^{l+1}\, ,
\\
\phi _n =-u_n(t)R^2/r\, +B_l(t)Y_l/r^{l+1}\, .
\end{gather*}
Using the same boundary conditions as above, we find the velocities of the unperturbed flow of the liquid
phase
\begin{eqnarray*}
 u_s=-\dot{R}\, (\rho '-\rho _s)/\rho _s\, , \;\;\;\;\;\;\;\; u_n=\dot{R}\, ,
\end{eqnarray*}
and coefficients $A_l(t)$ and $B_l(t)$ describing the perturbed motion of the interface
\begin{gather*}
A_l=\,\frac{\rho '-\rho _s}{\rho _s}\,\frac{R^{l+2}}{l+1}\left(\dot{\zeta}_l+\frac{2\dot{R}}{R}\,\zeta _l
\right) ,
\\
B_l=-\,\frac{R^{l+2}}{l+1}\left(\dot{\zeta}_l+\frac{2\dot{R}}{R}\,\zeta _l\right) .
\end{gather*}
Employing the same boundary relation for the pressures in the phases and the
same dependence between growth rate and chemical potential difference, we
obtain for the undisturbed growth of the solid phase after some algebra
\begin{equation}\nonumber
\rho _{\text{ef}}\!\left(\! R\ddot{R}+\frac{3}{2}\,\dot{R}^2\!\right)\! +\frac{\rho '}{K}\,\dot{R}=\frac{\rho
'-\rho}{\rho}\!\left(\!\Delta P-\frac{\rho}{\rho '-\rho}\,\frac{2\alpha}{R}\right) .
\end{equation}
The growth equation looks exactly like the motion of a particle with  the effective mass  $M(R)=4\pi\rho
_{\text{ef}}R^3$, drag force $4\pi R^2\rho 'K^{-1}\dot{R}$, and potential energy $U(R)=4\pi\alpha R^2-((\rho
'-\rho)/\rho )\Delta P(4\pi R^3/3)$, $\Delta P=P_{\infty}-P_c$ being overpressure.
\par
The small amplitude perturbations for the spherical surface of the solid phase are described by the relation
\begin{gather}\nonumber
\rho _{\text{ef}}\!\left(\frac{R}{l+1}\,\ddot{\zeta}_l+\frac{3}{l+1}\,\dot{R}\dot{\zeta}_l\right) +\frac{\rho
'}{K}\,\dot{\zeta}_l
\\
+\left(\alpha\,\frac{(l-1)(l+2)}{R^2}-\rho _{\text{ef}}\,\frac{l-1}{l+1}\,\ddot{R}\right)\zeta _l =0  .
\label{f4}
\end{gather}
For large radius $R$ and $l\rightarrow\infty$ so that $q=l/R$ is fixed,
equation~(\ref{f4}) goes over to the result for the plane geometry. Under
steady conditions $R(t)\equiv R$ the spectrum of crystallization waves on the
spherical interface is determined from
 $$ \omega _l^2 + i\omega _l\,\frac{\rho '}{\rho _{\text{ef}}K}\,\frac{l+1}{R}
 -\,\frac{\alpha}{\rho _{\text{ef}}R^3}\, (l-1)(l+1)(l+2) =0 .
 $$
Equation~(\ref{f4}) can have unstable solutions giving rise to the generation of crystallization waves at the
interface. The answer whether the spherical interface will be stable or unstable depends strongly on the
history of the time behavior $R(t)$. To illustrate, we consider the Richtmyer-Meshkov situation when a solid
with radius $R$ and interface rate $\dot{R}$ is subjected to a spherical shock acceleration
$\ddot{R}=V(0)\delta (t)$. In the linear regime the perturbation amplitude can be estimated according to
\begin{equation}\nonumber
\zeta _l(t)=\frac{V(0)\,(l-1)}{3\dot{R}+(l+1)\rho '\!/(\rho _{\text{ef}}K)\!}\!\left[1-\text{e}^{
-\left(\frac{3\dot{R}}{R}+\frac{\rho '(l+1)}{R\rho _{\text{ef}}K}\right)t}\right]\!\zeta _l(0) .
\end{equation}
Here $\zeta _l(0)$ is the initial perturbation amplitude of $l$-$th$ harmonic
at the interface immediately after the shock acceleration and $V(0)$ is an
additional velocity acquired by the interface. The mode with $l=1$ is
obviously not involved because this harmonic describes the displacement of a
sphere as a whole. Like the case of the planar geometry, the initial growth of
the interface perturbation is a linear function of time
 $$
\zeta _l(t)= tV(0)(l-1)\zeta _l(0)/R \, .
 $$
The long-time behavior of the interface perturbation is strongly governed by the magnitude and the sign of the
growth rate $\dot{R}$. In fact, provided that a solid $^4$He globe either grows at any rate or melts not so
fast at the moment of shock acceleration, i.e.,
 $$
\dot{R}> -\,\frac{1}{3}\frac{\rho '(l+1)}{\rho _{\text{ef}}K}\, ,
 $$
the growth of the perturbation amplitude saturates. For larger harmonics, the stabilizing role of the finite
damping of crystallization waves increases. In the opposite regime when a $^4$He solid melts sufficiently
fast, the linear growth of the interface amplitude in time will cross over into an exponential increment of
the perturbation amplitude.
\par
The effects in the spherical geometry can be studied by focusing a high-intensity sound wave in the middle of
an experimental cell. The experiments \cite{Is} on nucleation of solid $^4$He with two hemispherical
piezoelectric transducers glued together to make a spherical geometry have shown that it is possible to
achieve very high pressure amplitudes, more than 100~bar in the bulk liquid $^4$He.

\section{SUMMARY}
\par
We have investigated here the analogue of the Ray\-leigh-Taylor and Richtmyer-Meshkov instabilities of
crystallization waves at the accelerated superfluid-solid $^4$He interface. Our analysis, made within the
linear theory in perturbation, shows that the Rayleigh-Taylor and Richtmyer-Meshkov instabilities can occur as
well as the parametric Faraday instability \cite{Sa} which we have treated as a periodically driven version of
the Taylor instability. The plane and spherical interfaces are considered, and the first-order linearized
equations are found for the perturbation amplitudes.
\par
Unfortunately, it is difficult to make a well-founded conclusion in favor of destructing the crystal faceting
as a result of the impulsively accelerated interface. For a quantitative analysis, it is necessary to have the
images of crystal evolution taken successively in time. However, such aspects as the Taylor-like instability
cannot be rejected for the highly accelerated superfluid-solid $^4$He boundaries.
\par
Regarding the well-faceted and atomically smooth crystal surfaces which may have an infinitely large surface
stiffness, we can make the following remarks. The large value of surface stiffness $\gamma$ is a factor,
first, for nonzero wave vectors $q$, which prevents the development of the Taylor instability. The most
distinctive feature of the smooth crystal surface from the rough one is the existence of a non-analytic
cusp-like behavior in the angle dependence for the surface tension, e.g., \cite{BAP,No}. The presence of a
singularity leads to qualitative distinctions in the development of instabilities \cite{Bu}. First and
foremost, the threshold magnitude and the conditions determining the development of the Taylor instability
prove to be dependent on the initial amplitude of the interfacial perturbation. The smaller the initial
perturbation amplitude, the larger the necessary threshold magnitude of acceleration, approximately as
$1/\zeta _q(0)$. We note here that the onset of the Taylor instability is favored at the vicinal surfaces
whose orientations are tilted by a small angle with respect to the high-symmetry faceted ones. The larger the
slope of the vicinal plane, the smaller the initial amplitude of interfacial perturbations is required for the
development of the instability at the same magnitudes of acceleration.
\par
The constant acceleration of an interface works like an effective gravity. Hence, the physical picture for the
Taylor case of constant acceleration is analogous to the gravity-driven fingering of a crystal atop a fluid
and can qualitatively be interpreted in terms of an effective amplitude-dependent stiffness $\gamma \sim
1/q\zeta _q$ \cite{Bu}. For sufficiently large perturbation amplitudes $\zeta _q$, the difference between the
cases of smooth and rough crystal surfaces disappears.
\par
In the case of impulusively accelerated smooth interface, the very initial stage of instability will be similar
to that of a rough surface with the perturbation amplitude increasing linearly in time. Insensitivity to the
surface state results from the fact that the inertial properties of the interface, associated with its
effective density, are mainly responsible for a linear response on instant loading. However, the specific time
when the linear time growth of perturbations breaks down becomes amplitude-dependent for the smooth surfaces
and shorter as compared to the rough surfaces. \par Experimental and theoretical study of these effects can be
useful for clarifying physical aspects of the crystal $^4$He growth under high drives and fast dynamics.

\section{ACKNOWLEDGMENTS}
\par
We wish to acknowledge gratefully the helpful suggestions of R.~Barankov and assistance from R.~Hipolito. The
work is supported by the RFBR Grant No.~08-02-00752a.

\end{document}